\input{aipcheck}

\documentclass[
    ,final            
  ]
  {aipproc}

\layoutstyle{8x11single}

\newcommand{\pt}{p_T^\mu}

\begin{document}

\title{Muon Identification at ATLAS and CMS}


\author{Oliver Kortner}{
  address={Max-Planck-Institut f\"ur Physik, F\"ohringer Ring 6, D-80805
  M\"unchen, Germany}
}

%

\begin{abstract}
Muonic final states will provide clean signatures for many physics processes
at the LHC. The two LHC experiments ATLAS and CMS will 
be able to identify muons with a high reconstruction efficiency above 96\% 
and a high transverse momentum resolution better than 2\% for transverse 
momenta below 400 GeV/c and about 10\% at 1 TeV/c. 
The two experiments follow complentary concepts of muon detection. ATLAS has an 
instrumented air-toroid mangetic system serving as a stand-alone muon 
spectrometer. CMS relies on high bending power and momentum resolution in 
the inner detector, and uses an iron yoke to increase its magnetic field. 
The iron yoke is instrumented with chambers used for muon identification. 
Therefore, muon momenta can
only be reconstructed with high precision by combining inner-detector
information with the data from the muon chambers.
\end{abstract}

\maketitle


\section{Introduction}
The large hadron collider (LHC) will be the next hadron collider which will go
into operation. From the year 2008 on, it will collide protons at a 
centre-of-mass energy of 14~TeV with
a luminosity between 10$^{33}$~cm$^{-2}$\,s$^{-1}$ and 
10$^{34}$~cm$^{-2}$\,s$^{-1}$ \cite{LHC}. Many physics processes which are
currently out of reach at the existing colliders will become accessible by the
LHC. These processes range from
the production of the Higgs and new gauge bosons to the production of
superpartners of the known fundamental particles. All of them, however, 
are highly obscured by QCD reactions. The QCD reactions are usually
accompanied by multiple jets in the final state and leptons of low
transverse momenta. Therefore many of the interesting non-QCD
processes can be detected in final states with leptons of high transverse
momenta. Since muons lose very little energy on their passage through matter
and are therefore the only highly energetic
charged particles traversing the entire
detector, muonic final states provide the cleanest signatures for the
detection of new physics processes.

\begin{figure}[hbt]
	\includegraphics[width=0.5\linewidth]{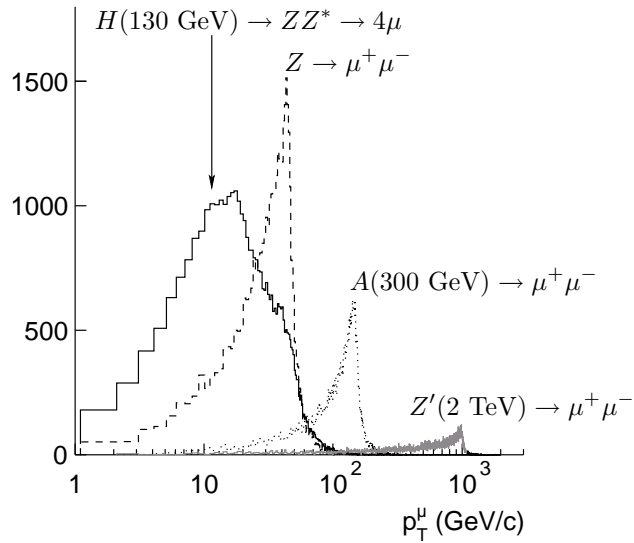}
	\caption{Transverse momentum distribution for muons from $H$, $Z$, $A$,
	and $Z'$ decays as predicted by Pythia \cite{Pythia}.}
	\label{kortner_fig1}
\end{figure}
Figure~\ref{kortner_fig1} shows the transverse-momentum distributions of muons 
from the decays of $Z$ bosons, standard-model Higgs bosons $H$ with a mass of
130~GeV$/$c$^2$,
supersymmetric Higgs bosons $A$ with a mass of 300~GeV$/$c$^2$, and $Z'$ bosons 
with a mass of 2~TeV$/$c$^2$. As the masses of the resonances cover a range from
90~GeV$/$c$^2$ to 2~TeV$/$c$^2$, the decay muons have transverse momenta from a
few GeV$/$c to 1~TeV$/$c. As a consequence, the detectors at the LHC must be
able to detect and identify muons over this large transverse momentum range.

\begin{figure}[hbt]
	\includegraphics[width=0.5\linewidth]{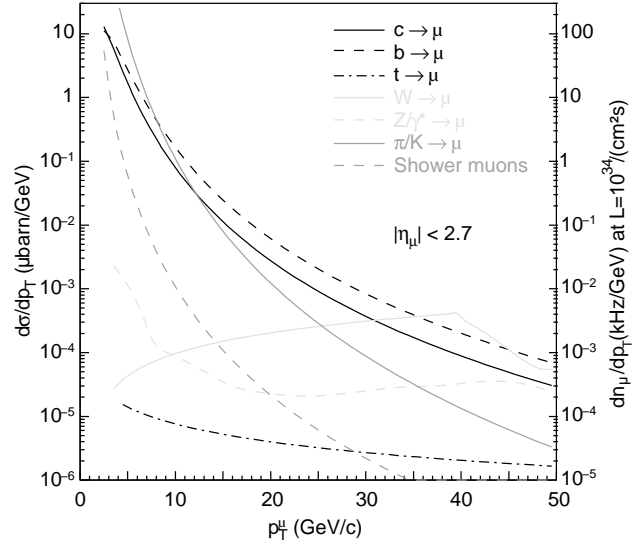}
	\caption{Transverse momentum dependence of inclusive muon cross sections
	for the ATLAS experiment at the LHC \cite{ATLAS_mTDR}.}
	\label{kortner_fig2}
\end{figure}
Yet the main source of muons are mesons as can be concluded from the 
inclusive muon cross sections in Figure~\ref{kortner_fig2}. The cross sections
presented there depend on the hermicity and density of
the calorimeters of the detector; the detectors under discussion in this
article have similar cross sections. Most of the muons with
transverse momenta $\pt<$10~GeV$/$c are from the decays of charged pions and 
kaons in the final state of the initial proton-proton collision.
At $\pt\approx$10~GeV$/$c, decays of $b$ and $c$ mesons become the dominant 
source
of muons. The inclusive cross sections for muons from $W$ and $Z$ decays are 
only comparable to the inclusive cross sections for muons from $b$ and $c$ 
meson decays in the region around 50~GeV$/$c. 
Muons from $t$ quark decays are much rarer than from $b$ and $c$ decays.
Charged pions are not only produced in the primary proton-proton collision but
also during the absorption of the final state hadrons in the calorimeters. The
pions in hadron showers, however, have lower transverse momenta than the
primarily produced pions. Hence the shower muons, i.e. the muons from
the decays of pions in hadron showers, are very soft so that most of them cannot
escape the calorimeter and remain undetected. 
At the LHC, shower muons are insignificant as compared to muons from the decay
of pions and kaons before their interaction in the detectors.
Although the muons of all the different sources can be detected as muons by the
particle detectors at the LHC, it has become common practice in muon
identification to identify only those muons as muons which do not emerge from
pion and kaon days or hadron showers.

Two omnipurpose experiments will be operated at the LHC: ATLAS (a toroidal LHC
apparatus) and CMS (the compact muon solenoid). Both experiments have a high
muon detection and identification capability as will be explained in the 
present article.

\section{The ATLAS and CMS Muon Systems}
In the ATLAS and CMS detectors, muons are identified as those charged particles
which
leave a trace in the inner tracker which matches a charged track in
tracking chambers mounted outside the calorimeters. Muons lose about 3~GeV of
their energy on average
during their passage through the calorimeters. The energy loss of
muons is dominated by ionization up to muon energies of 100~GeV. As the energy
lost in a single ionization is small, the overall energy loss has a small spread
about its mean value. 
Above 100~GeV, direct e$^+$e$^-$ pair production, bremsstrahlung, and
nuclear losses contribute significantly to the energy loss of muons 
\cite{Zupancic}. The latter two processes may relatively often lead to a single
"catastrophic" energy loss comparable to the total muon energy. This causes a
low-energy tail in the measured muon energy. Its spectrum still exhibits a peak
with a $\sigma$ of about 20\% for $E_\mu=1~$~TeV. 
The measured muon energies in the tail of the distribution can be
recovered by means of the energy detected along the muon trajectory
in the calorimeter.

The energy of a muon at the exit of 
the calorimeters is therefore equal to the energy of the muon plus its well
calculable energy loss in the calorimeters with a relative accuracy of about 1\%
for muon energies above 10~GeV.

\begin{figure}[hbt]
	\includegraphics[width=\linewidth]{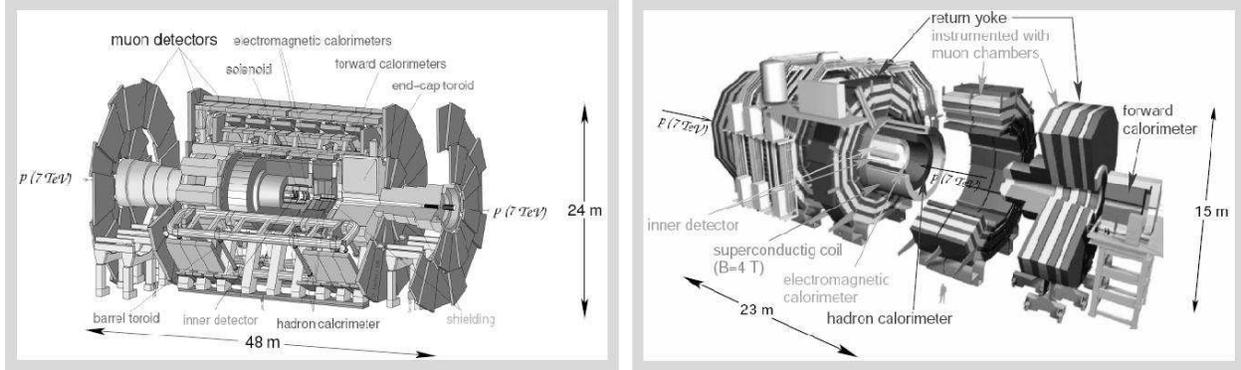}
	\caption{Sketch of the ATLAS detector on the left and the CMS detector
	on the right.}
	\label{kortner_fig3}
\end{figure}

Both ATLAS and CMS have the capability of measuring the energies of the muons in
the muon systems. They differ, however, in accuracy as
the designs of the detectors follow different strategies. Sketches of the ATLAS
and CMS detectors are shown in Figure~\ref{kortner_fig3}. Both detectors have
the typical structure of collider detectors: an inner tracking detector in a
solenoidal magnetic field surrounded by electromagnetic and hadron calorimeters
which are themselves enclosed in a muons system.

The muon system of ATLAS \cite{ATLAS_mTDR} 
is designed to be able to measure muon momenta with
high accuracy independently of the inner detector. It uses a
system of eight superconducting air-core toroid coils, which produces a toroidal
magnetic field of 0.4~T on average in the muon system. The air-core structure
was used to minimize the amount of material the muons have to traverse
in the muon spectrometer, hence to minimize multiple scattering of muons in the
muon spectrometer. The deflections of the muons in the barrel part of the
spectrometer are measured by three layers of tracking chambers with 2.5~m
interspace. The inner and outer layers are attached to the toroid coils, 
the middle layer is mounted in the middle of the coils. The endcap part of the muon chambers
consists of three disks of tracking chambers at 6~m interspace with the two
inner disks enclosing the endcap toroid system (see Figure~\ref{kortner_fig3}). 
The sagittae of curved muon
trajectories in the spectrometer are of the order of
$\frac{0.5~\mathrm{m}}{p^\mu[\mathrm{GeV/c}]}$, thus 0.5~mm for
$p^\mu=1$~TeV$/$c. In order to measure the sagitta of 0.5~mm for 1~TeV muons
with an accuracy of 10\%, ATLAS uses monitored drift-tube chambers which have a 
spatial resolution better than 0.040~mm. An optical alignment system is used to
monitor the geometry of the muon spectrometer with 30~$\mu$m accuracy.
Since the monitored drift-tube chambers
have a maximum response time of 700~ns and are therefore slow compared to the
25~ns bunch-crossing frequency of the LHC, resistive-plate chambers in the
barrel part and thin-gap chambers in the endcap part of the muon spectrometer
are used for triggering. The trigger chambers have a response time below 10~ns
and are therefore capable of assigning the measured muons to the correct
bunch crossing. The muon spectrometer is capable of detecting muons up to
pseudorapidities of 2.7.

In the design of the CMS detector great emphasis
is given on a very high
momentum resolution of the inner detector. The momentum
resolution is inversely proportional to the bending power inside the inner 
detector.
A magnetic field strength of 4~T in the inner detector is achieved by amplifying
the magnetic field with a large iron yoke embedding the superconducting solenoid
coil. For comparison: as ATLAS abstains from an iron yoke, the magnetic field in 
the inner detector of ATLAS is only 2~T. The return yoke of the CMS solenoid is
instrumented with four layers of muon chambers. As the momentum resolution of the
CMS muon system is dominated by multiple scattering of the muons in the iron 
yoke, tracking chambers with a lower spatial resolution ($\approx$70~$\mu$m) 
than 
in ATLAS can be used. Drift-tube chambers with 400~ns response time are used in
the barrel part of the CMS muon system, cathode strip chambers with 50~ns
response time in the endcaps. Like in ATLAS, fast trigger chambers are 
needed to identify the detected muons with the right bunch crossing of the LHC.
Resistive-plate chambers are used throughout the whole CMS muon system for
triggering. The CMS muon system has a pseudorapidity coverage of 2.4.

\begin{figure}[hbt]
	\includegraphics[width=\linewidth]{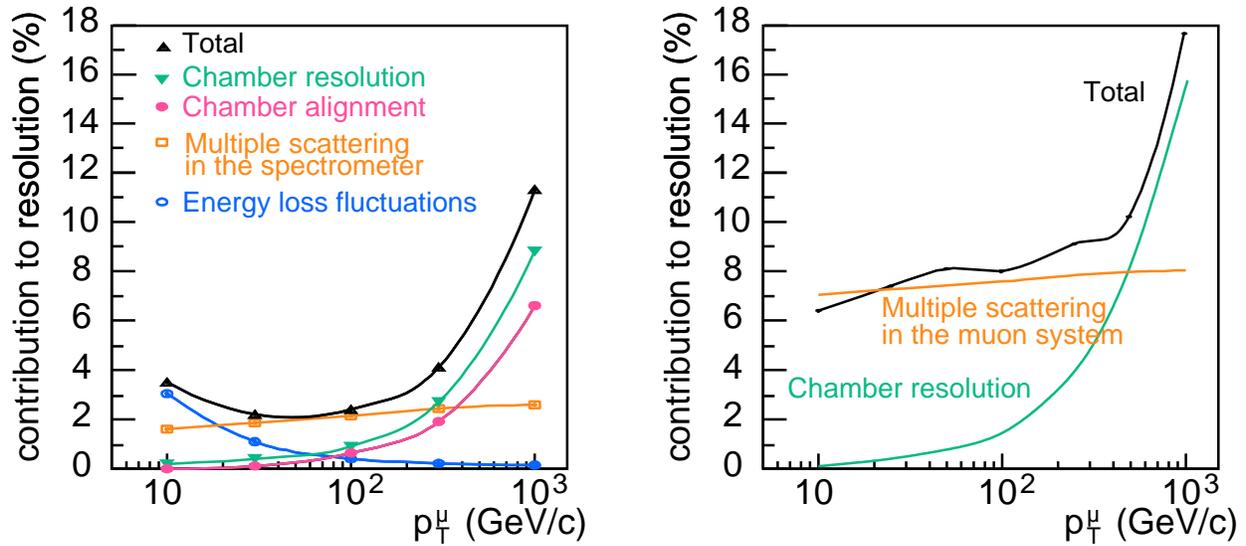}
	\caption{Momentum resolution of the barrel
	muon system in stand-alone mode.
	Left: ATLAS \cite{ATLAS_mTDR}. Right: CMS \cite{CMS_mTDR} (the
	contributions to the resolution are estimates by the author of this
	article).}
	\label{kortner_fig4}
\end{figure}

The muon momentum resolutions which can be achieved in stand-alone mode of the
ATLAS and CMS barrel muon systems are shown in Figure~\ref{kortner_fig4}. 
The ATLAS muon spectrometer has a high transverse momentum resolution between 2 and 
4\% for transverse muon momenta $\pt\le300$~GeV$/$c. The $\pt$ resolution
reaches 12\% for $\pt=1$~TeV$/$c. Energy loss fluctuations
only influence the momentum resolution at very low $\pt$. The resolution
has a lower limit of 2\% caused by multiple scattering in the muon spectrometer.
For $\pt>300$~GeV$/$c the curvatures of the muon trajectories in the muon
spectrometer are so small that the intrinsic spatial resolution of the
chambers and the limited accuracy of the chamber alignment dominate the momentum
resolution of the spectrometer. In the CMS muon system, the transverse momentum
resolution is limited by the chambers resolution for high values of $\pt$, but
multiple scattering of
the muons in the iron yoke is the dominant contribution to its transverse
momentum resolution up to $\pt=$400~GeV$/$c. The transverse momentum
resolution is proportional to the amount of multiple scattering and inversely 
proportional to the bending power of the magnetic field in the muon system. The
amount of multiple scattering is about 14 times larger in CMS than in ATLAS, but
the bending power of the magnetic field is 5 times larger in CMS than in ATLAS
such that, overall, the stand-alone transverse momentum resolution is only 3
times worse than in ATLAS.

\begin{figure}[hbt]
	\includegraphics[width=\linewidth]{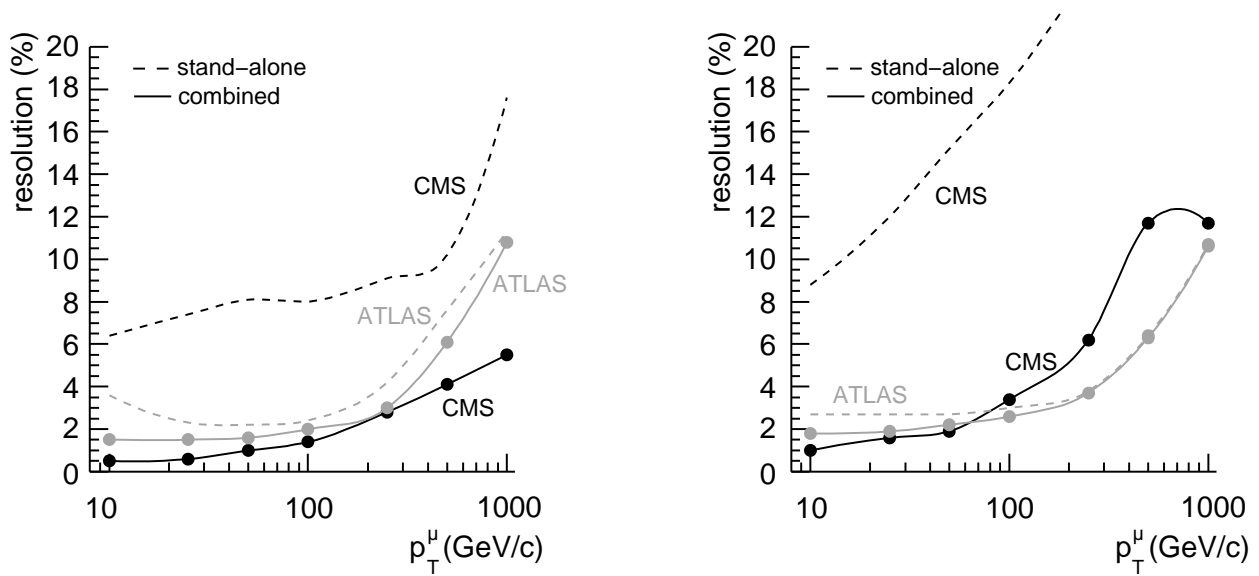}
	\caption{Momentum resolution of the ATLAS and CMS detector after
	combining the momentum measurements of the inner detectors and the muon
	systems \cite{ATLAS_mTDR}\cite{CMS_mTDR}.
	Left: Momentum resolution for muons in the barrel region. 
	Right: Momentum resolution for muons in the endcap region.}
	\label{kortner_fig5}
\end{figure}

The situation improves significantly after the combination of the momentum
measured by the muon system with the momentum measured by the inner detector
(see Figure~\ref{kortner_fig5}). CMS achieves an overall $\pt$ resolution
between 0.5 and 2\% for $\pt\le300$~GeV$/$c due to the high bending power and
the high resolution of the inner detector system. The bending power in the ATLAS
inner detector is about half as big as in CMS and the combined muon momentum
resolution is about twice poorer than in CMS. The bending power of the CMS
magnet is much smaller in the endcaps than in the barrel part of the muon system
which makes the stand-alone muon momentum resolution worse. The measurement of
the momentum of muon flying into an endcap is totally determined by the inner
detector. ATLAS profits from its high stand-alone performance and achieves a
better combined momentum resolution than CMS for high $\pt$ muons in the 
endcaps.

\section{Muon Track-Reconstruction and Identification}
All high momentum tracks which are recorded by the muon system of ATLAS and CMS
are muon tracks. Both experiments follow the same track reconstruction
strategy. 

\begin{figure}[hbt]
	\includegraphics[width=0.4\linewidth]{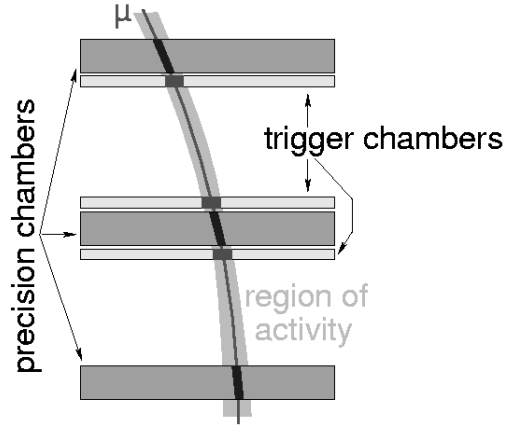}
	\caption{Track reconstruction in the barrel of the ATLAS muon
	spectrometer.}
	\label{kortner_fig6}
\end{figure}

Figure~\ref{kortner_fig6} illustrates the track reconstruction steps in
case of a muon with $\pt>10$~GeV$/$C
in the barrel part of the ATLAS muon spectrometer.
As the trigger chambers have a much lower granularity and simpler geometry than
the precision chambers, their hits can easily be used to define a so-called
"region of activity" through which the muon must have passed. In the second step
of the track reconstruction, straight segments are reconstructed in the
precision chambers within the region of activity. The local straight segments 
of the region of activity are then combined to a curved candidate
trajectory which is finally refined by a global refit of the hits lying on 
the candidate trajectory.

The combination of the trajectory found in the muon system with a matching
trajectory in the inner detector serves two goals: (1) the improvement of the
momentum resolution, especially in case of CMS where the stand-alone muon
momentum resolution is poor compared to the momentum resolution of the inner
detector; (2) the rejection of muons from pion and kaon decays. The mother pions
and kaons of the muons have decays lengths of a few metres. Thus most of them 
leave a long trace in the inner detector before they decay. Since considerable
parts of the energies of the mother pions or kaons are carried away by
neutrinos, the energies of the decay muons is significantly lower than the
energies of their mothers and the momentum of the inner-detector track is in
mismatch with the muon-system track. Therefore muons from pion or kaon decays
can be rejected by requiring a good match of the corresponding
inner-detector and muon-spectrometer trajectories.

The procedure described above cannot be applied to muons of
$\pt<10$~GeV$/$c because their trajectories do not extend to the outermost
layers of the muons systems. In case of CMS, they are absorbed by the iron yoke
before; in case of ATLAS, the magnetic field in the spectrometer is strong
enough to bend them away from them. So the reconstruction starts with a
low-momentum inner-detector track in the same solid angle as segments found in
the muon system, extrapolates it to the muon system,
and the segments lying on the extrapolation are selected. The number of these
segments, their length and position must be compatible with the momentum of the
inner-detector track. This requirement rejects muons from pion and kaon decays.
Pions and kaon decaying in the calorimeters deposit much more energy in the
calorimeters than muons which are minimum ionizing particles. One therefore also
requires that the energy deposited along the extrapolated inner detector
trajectory in the calorimeters is compatible with the muon hypothesis.
Monte-Carlo studies \cite{CMS_tracking}\cite{MUTAG} show that an efficiency
$>80$\% and a fake rate $<0.5$\% can be achieved for muons of $\pt>5$~GeV$/$c in
both experiments.

\begin{figure}[hbt]
	\includegraphics[width=0.5\linewidth]{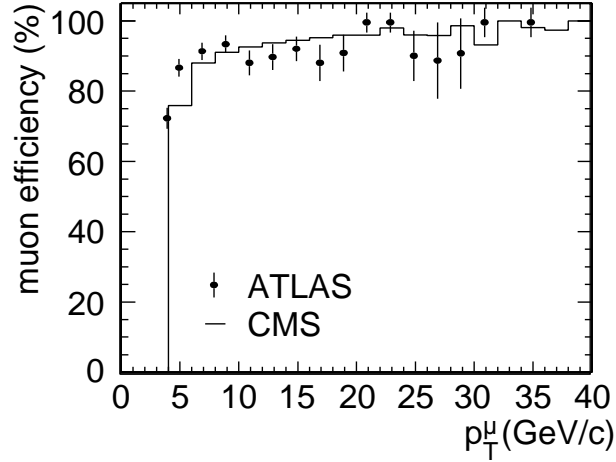}
	\caption{Muon identification efficiency of ATLAS\cite{MUTAG} and CMS
	\cite{CMS_tracking}.}
	\label{kortner_fig7}
\end{figure}

Figure~\ref{kortner_fig7} summarizes the muon identification efficiency of
ATLAS and CMS. The two experiments achieve the same track-reconstruction
efficiency, namely about 90\% for $5<\pt<20$~GeV$/$c and $>96$\% for
$\pt>20$~GeV$/$c at a fake rate $<0.5$\%.

This result is particularly difficult to achieve in case of ATLAS. Simulations
\cite{high_background} show that, different from CMS, a sizeable amount of 
low-energy neutrons leaks out of the calorimeters into the muon spectrometer.
These neutrons excite nuclei in the entire experimental hall of ATLAS such that
the muon spectrometer is operated in a huge neutron and $\gamma$ background.
This background leads to occupancies of up to 20\% in the precision chambers,
but less than 0.3\% in the trigger chambers due to their much shorter response
time. The low occupancy of the trigger chambers is responsible for a reliable
determination of the regions of activity in the track reconstruction in the muon
spectrometer. Test-beam studies \cite{X5_Elba}\cite{IEEE}
and Monte-Carlo simulations \cite{ATLAS_mTDR} show that efficient track
reconstruction in the precision chambers is possible under these circumstances.

\section{Summary}
Leptonic and especially muonic final states are best suited for physics
discoveries at the LHC. The two omnipurpose experiments ATLAS and CMS which
will be operated at the LHC are capable of detecting and identifying muons with
an efficiency $>90\%$ and a fake rate $<0.5\%$ over a wide muon momentum
interval of $5<\pt<1$~TeV$/c$. Both experiments achieve a $\pt$ resolution
$<$12\% over the whole momentum range.


\begin{theacknowledgments}
The author would like to express his thanks to Eric Lancone, Stefane Willoc, 
and Kerstin M\"uller for providing him information about the muon 
reconstruction and identification strategies in ATLAS and CMS.
\end{theacknowledgments}



\bibliographystyle{aipproc}   

\bibliography{kortner}

\IfFileExists{\jobname.bbl}{}
 {\typeout{}
  \typeout{******************************************}
  \typeout{** Please run "bibtex \jobname" to optain}
  \typeout{** the bibliography and then re-run LaTeX}
  \typeout{** twice to fix the references!}
  \typeout{******************************************}
  \typeout{}
 }

\end{document}